\documentstyle[sprocl,epsfig]{article}
\def\ee{$e^+e^-$}               

\def\nbar{\bar n}               

\def\NF{\mathcal{N}_{\kern -1.9pt f}}     
\def\NC{\mathcal{N}_{\kern -1.7pt c}}     

\newcommand{\roots}[1]{$\sqrt{s} = #1$ GeV} 
\def\pt{p\kern -.2pt\lower 4pt\hbox{\tiny T}}    

\def\p0{P_0(\Delta y)}

\def\ymin{y_{\mathrm{min}}}
\def\NB{{\mathrm{NBD}}}
\def\chiDF{$\scriptstyle \chi^2$/{\small df}}

\begin{document}
\thispagestyle{empty}
\begin{center}
\hfill DFTT 56/96\\
\hfill MPI-PhT/96-90\\
\hfill LU TP 96-25\\
\vspace{0.7cm}

{\bf COMMON ORIGIN OF THE SHOULDER STRUCTURE AND OF THE OSCILLATIONS
OF MOMENTS IN MULTIPLICITY DISTRIBUTIONS IN $e^+e^-$ ANNIHILATIONS}
\vspace{0.8cm}

        A. Giovannini \footnote[1]{E-mail: giovannini@to.infn.it}\\
{\it Dip.\ di Fisica Teorica and I.N.F.N. - Sezione di Torino, \\
  via P. Giuria 1, I-10125 Torino, Italy}\\
\vspace{0.2cm}
        S. Lupia \footnote[2]{E-mail: lupia@mppmu.mpg.de}\\
{\it Max-Planck-Institut f\"ur Physik (Werner-Heisenberg-Institut)\\
    F\"ohringer Ring 6, D-80805 M\"unchen, Germany}\\
\vspace{0.2cm}
        R. Ugoccioni \footnote[3]{E-mail: roberto@thep.lu.se}\\
{\it Department of Theoretical Physics, University of Lund,\\
   S\"olvegatan 14A, S-22362 Lund, Sweden}

\vspace{0.8cm}

ABSTRACT\\
\end{center}

\noindent We show, 
via a simple parametrization of the multiplicity distribution of
charged particles in $e^+e^-$ annihilation at the $Z^0$ peak 
in terms of the weighted
superposition of two negative binomial distributions, that both the shoulder
structure in the intermediate multiplicity range and the oscillation in sign of
the ratio of factorial cumulants over factorial moments of increasing order are
related to hard gluon radiation.

\begin{center}
\vspace{0.8cm}
Talk presented at the\\ 7th International Workshop ``Correlations and
Fluctuation''\\ (Nijmegen, The Netherlands, June 30 -- July 6, 1996)
and at the\\ 28th International Conference on High Energy Physics --
ICHEP 96\\ (Warsaw, Poland, July 25--31, 1996).
\end{center}

\newpage

\title{COMMON ORIGIN OF THE SHOULDER STRUCTURE AND OF THE OSCILLATIONS
OF MOMENTS IN MULTIPLICITY DISTRIBUTIONS IN \ee\ ANNIHILATIONS}

\author{A. Giovannini}
\address{Dip.\ di Fisica Teorica and I.N.F.N. - Sezione di Torino, \\
  via P. Giuria 1, I-10125 Torino, Italy}
\author{S. Lupia}
\address{Max-Planck-Institut f\"ur Physik (Werner-Heisenberg-Institut)\\
    F\"ohringer Ring 6, D-80805 M\"unchen, Germany}
\author{R. Ugoccioni}
\address{Department of Theoretical Physics, University of Lund,\\
   S\"olvegatan 14A, S-22362 Lund, Sweden}
\maketitle

\abstracts{We show, 
via a simple parametrization of the multiplicity distribution of
charged particles in \ee\ annihilation at the $Z^0$ peak 
in terms of the weighted
superposition of two negative binomial distributions, that both the shoulder
structure in the intermediate multiplicity range and the oscillation in sign of
the ratio of factorial cumulants over factorial moments of increasing order are
related to hard gluon radiation.}

\section{Common origin of shoulder structure and moments' oscillations}

One of the main still open problems in multiparticle dynamics is
that of attaining an integrated description of final particle Multiplicity
Distributions (MD's) and of the corresponding correlation functions properties.
One expects that features detected  in terms of one of the two observables
have a sound  physical counterpart in terms of the other. It is of course a 
quite difficult task to explain  facts occurring in the two domains 
by means of the same physical cause and accordingly to show that they have 
a common origin.
We discuss in the following a successful
example of this search in \ee\ annihilation. Here  two interesting
features are observed  at the $Z^0$ energy:
the MD shows a shoulder in the intermediate multiplicity 
range~\cite{delphi:2,opal,aleph,sld}
and the ratio of factorial moments to factorial cumulant moments
changes sign as a function of its order.\cite{sld,gianini}
The two observables are strictly linked, as 
factorial moments, $F_q$, and
factorial cumulant moments, $K_q$,  can be obtained in general from the MD,
$P_n$, through the relations:
\begin{equation}
  F_q = \sum_{n=q}^{\infty} n(n-1)\dots(n-q+1) P_n  	\label{eq:facmom}
\end{equation}
and
\begin{equation}
  K_q = F_q - \sum_{i=1}^{q-1} {q-1 \choose i} K_{q-i} F_i .  \label{eq:faccum}
\end{equation}
The ratio of the two mentioned quantities:
\begin{equation}
  H_q = K_q / F_q					\label{eq:hq}
\end{equation}
is well suited to theoretical and experimental studies.\cite{dremin-nech}

\begin{figure}[t]
\begin{center}
\mbox{\epsfig{file=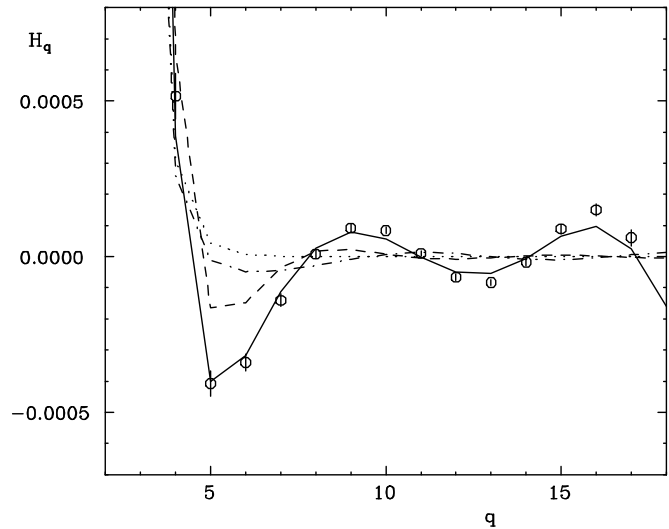,bbllx=86pt,bblly=286pt,bburx=473pt,%
bbury=596pt,width=9cm}}
\end{center}
\caption[Compare parameterizations]{
The ratio of factorial cumulant moments over factorial moments,
$H_q$ as a function of $q$;
experimental data (diamonds) from the SLD Collaboration~\cite{sld}
are compared with the predictions of several parameterizations, 
with parameters fitted to the data on MD's: a full NBD (dotted line); 
a truncated NBD (dot-dashed line); 
sum of two full NBD's as per Eq.~\ref{eq:formula} (dashed line); 
sum of two truncated NBD's as per Eq.~\ref{eq:truncformula} (solid line).
}\label{fig:all4}
\end{figure}

QCD calculations are available for both the MD~\cite{dokshitzer} 
and the ratio $H_q$.\cite{dremin-nech}
Both fail to
reproduce the data after a simple application of Local Parton Hadron
Duality~\cite{LPHD} as hadronization prescription
(or its generalized version,\cite{AGLVH:2}
to be used with moments of higher order~\cite{VietriRU}). 
The most recent calculation of the MD has been done in MLLA 
gluodynamics:\cite{dokshitzer} KNO scaling is 
present only asymptotically,
and the predicted shape at current energies is much narrower than in
previous calculations,\cite{DKMT} but still
the head and the tail of the distribution cannot be correctly reproduced. 
The ratio $H_q$ has also been calculated beyond DLA,\cite{dremin-nech} and
has been predicted to show a negative minimum around $q=5$ and then to
oscillate in sign as a function of the order q; this agrees qualitatively with
the behaviour seen in the data.\cite{gianini}

Many phenomenological parameterizations have been used to describe data
on MD's, the most common (and simplest) being the Negative Binomial Distribution
(NBD)~\cite{AGLVH:1} and the 
Log-Normal Distribution (LND).\cite{LND} 
Both work well at lower energies, but fail to reproduce the shoulder
structure at \roots{91}.
In particular this failure can be seen if the residuals (difference
between the data and the parametrization divided by the error on the data)
are examined: a clear structure is seen.\cite{sld}
The SLD Collaboration has also shown that these parameterizations fail
in the sector of the $H_q$ moments,\cite{sld} even after taking 
into account the effect of finite statistics.\cite{hqlett}
This is particularly evident in Figure~\ref{fig:all4}, if one looks at 
the dotted (full NBD) and dot-dashed (truncated NBD) lines.

The shoulder structure was first explained by the Delphi
Collaboration:~\cite{delphi:2}
it results from the superposition of samples of
events with a fixed number of jets. In each sample taken separately
there is no shoulder but, since they have a different
average multiplicity, a shoulder appears in the full sample.
In addition it was found that the NBD describes well the MD in these 
samples of events
with a fixed number of jets.\cite{delphi:3}
It should also be remembered that a shoulder has been seen
in $p\bar p$ collisions at \roots{900} by the UA5 Collaboration,\cite{UA5:3}
and it has
been confirmed at the Tevatron.\cite{rimondi}
A good fit to the UA5 data was performed
with the weighted sum of two NBD's.\cite{fuglesang}

Following the above observations,
we propose a parametrization of the MD which is the weighted sum
of two components, one to be associated with 2-jet events and
one to be associated with events with 3 or more jets. The weight
in this superposition is then the fraction of 2-jet events,
which is experimentally determined, it is not a fitted parameter.
This decomposition depends of course on the definition of jet; in
particular it depends on the particular jet-finding algorithm, and
on the value of the parameter $\ymin$ which controls the algorithm. The
Delphi Collaboration has used the JADE algorithm and published
values for the 2-jet fraction and for the MD's at $\ymin$ = 0.02, 0.04,
0.06, 0.08. We have performed the fit for each of these values.

Concerning the experimental data,
it should be noticed that the experimental errors on each point of the
published MD's are correlated with adjacent bins:
in our fit we cannot take this correlation into account,
therefore it is not possible to compare directly the values of the
$\chi^2$ we obtain with those obtained by the experimental
collaborations. Furthermore, the extraction of $H_q$ from the published
$P_n$ also suffers from a similar problem, and the errors we show
in the figures are estimates obtained by a statistical
method~\cite{hqlett:2} (except for the data
of the SLD Collaboration~\cite{sld}).

As for the particular form of the MD in the two components of
our fit, we have chosen the NBD, because it has successfully been
fitted to the data for the samples of events with fixed number of jets.
In practice we perform a fit to the MD's with a four parameter
formula:
\begin{equation}
  P_n \propto \cases{ \alpha P_n^{\NB}(\nbar_1,k_1) + (1-\alpha)
	P_n^{\NB}(\nbar_2,k_2) & \parbox{3.5cm}{if 
         $n$ is even}\cr
  0 & otherwise\cr }				\label{eq:formula}
\end{equation}
Here $P_n^{\NB}(\nbar,k)$ is the standard NBD of parameters $\nbar$ and $k$;
notice that we have taken into account the charge conservation law, which
requires the final charged particle multiplicity to be even. The
proportionality factor is fixed by requiring the proper normalization 
for $P_n$.

\begin{figure}
\begin{center}
\mbox{\epsfig{file=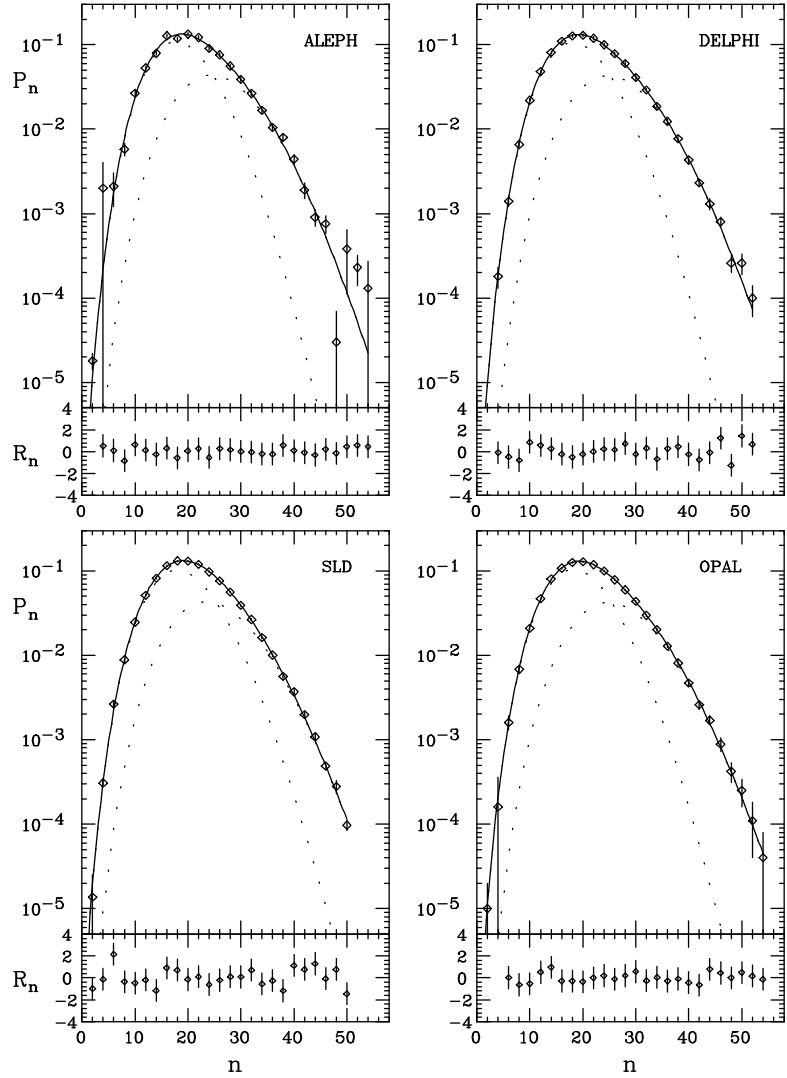,bbllx=42pt,bblly=71pt,bburx=524pt,%
bbury=735pt,height=14.5cm}}
\end{center}
\caption[fit to the MD]{
Charged particles MD's in full phase space, $P_n$, at the $Z_0$ peak  from
ALEPH,\cite{aleph} DELPHI,\cite{delphi:2} SLD~\cite{sld} and
OPAL~\cite{opal} Collaborations are
compared with Eq.~\ref{eq:formula}
with $\alpha$ = 0.767 (see Table~\ref{table} for the values of the
corresponding parameters) (solid lines);
dotted lines indicate the two separate NBD contributions.
The lower part of each plot shows the residuals, $R_n$, i.e., the difference
between data and theoretical predictions, in units of 
standard deviations.}\label{fig:fit}
\end{figure}

Results of our fit to the data of four experiments are shown in
Table~\ref{table}:
we find $\chi^2$ per degree of freedom equal to or smaller than 1, 
and values of the parameters consistent between
different experiments; 
they are also consistent with those obtained
by the Delphi Collaboration 
in fitting their 2-jet and 3-jet data separately with
a NBD. These findings are visually summarized 
for $\alpha=0.767$ in Figure~\ref{fig:fit},
where the residuals do not show structures.

\begin{figure}
\begin{center}
\mbox{\epsfig{file=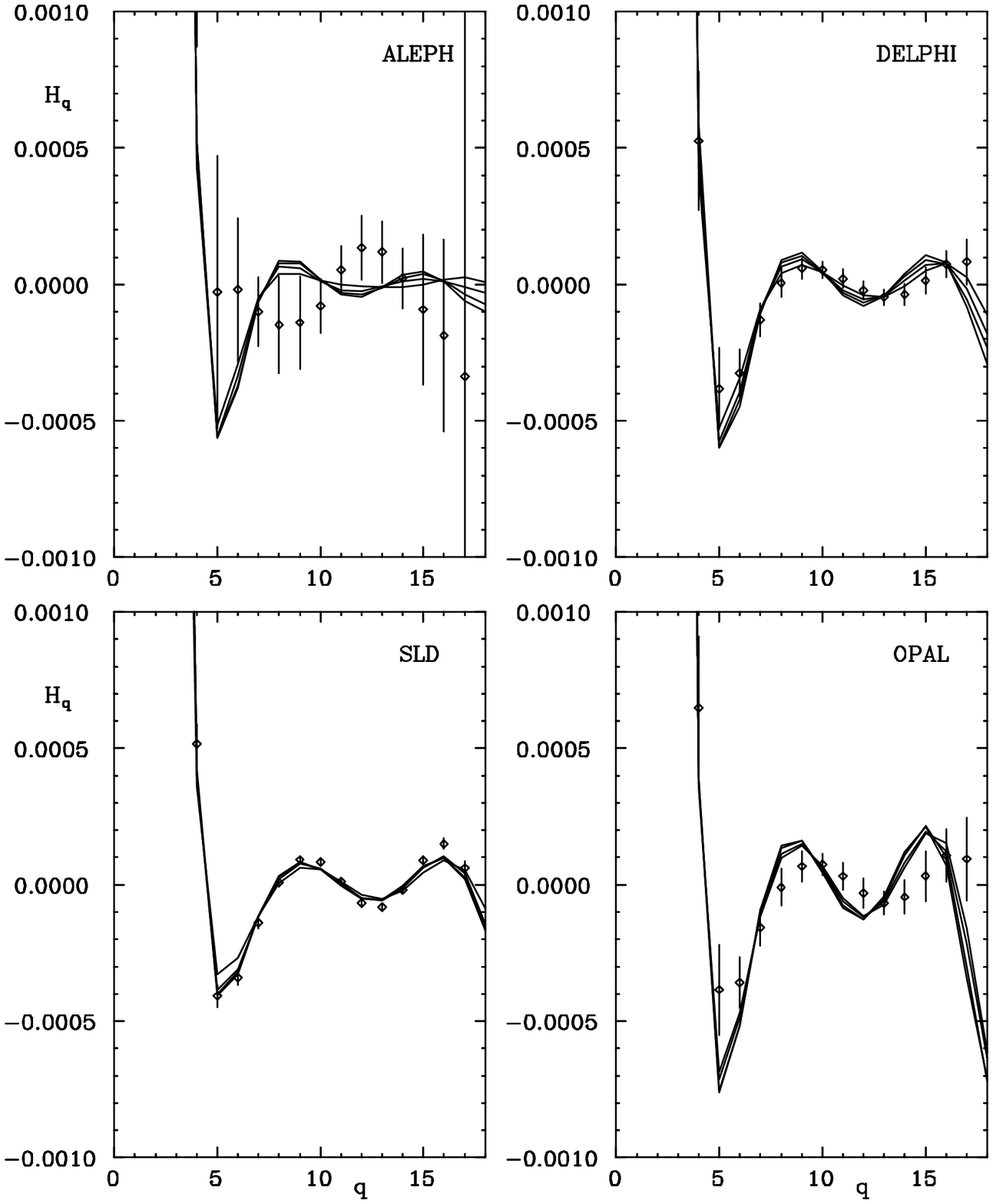,bbllx=34pt,bblly=111pt,bburx=529pt,%
bbury=705pt,width=11.8cm}}
\end{center}
\caption[Hq moments]{The ratio of 
factorial cumulant moments over factorial moments,
$H_q$ as a function of $q$;
experimental data (diamonds) from ALEPH, DELPHI, SLD  and OPAL 
Collaborations
are compared  with Eq.~\ref{eq:truncformula}
for different values of $\alpha$ (see Table~1 for the values
of the corresponding parameters).
In the figure only statistical errors of SLD data~\cite{sld}
are shown.
}\label{fig:hq}
\end{figure}

In Figure \ref{fig:hq} we compare the experimental data on $H_q$'s
with the values obtained from a formula that takes the
truncation effect into account, too:
\begin{equation}
  \tilde P_n \propto \cases{ P_n & \parbox{3.5cm}{if 
        ($n_{\mathrm{min}} \le n \le n_{\mathrm{max}}$)}\cr
  0 & otherwise\cr }				\label{eq:truncformula}  
\end{equation}
where $n_{\mathrm{min}}$ and $n_{\mathrm{max}}$ are the minimum and
maximum observed multiplicity, and a proportionality factor ensures
proper normalization. We obtain a very good agreement with the data.
Notice that 
it is not possible to
reproduce the behaviour of the ratio $H_q$ without taking into account the
limits of the range of the  available data. 
This can be seen
in Figure~\ref{fig:all4}, by comparing the dashed (full formula)
and the solid (truncated formula) lines.

In conclusion, the observed behavior of $H_q$'s results from the
convolution of two different effects, a statistical one, i.e., the
truncation of the tail due to the finite statistics of data samples, and 
a physical one, i.e., the
superposition of two components.
The two components can be related to 2- and 3-jet
events, i.e., to the emission of hard gluon radiation in the early stages of
the perturbative evolution.

\begin{table}
\caption[Parameters of the fit]{Parameters and 
$\chi^2$ per Degree of Freedom (df) of the fit to experimental data from
ALEPH,\cite{aleph} DELPHI,\cite{delphi:2} OPAL~\cite{opal} and SLD~\cite{sld}
Collaborations with the weighted sum of two NBD's.
Results are shown for different values of $\alpha$ corresponding to the
fraction of 2-jet events, $f$, experimentally measured by DELPHI 
Collaboration~\cite{delphi:3}
  at different values of the jet-finder parameter
$\ymin$. NBD parameters extracted by the DELPHI Collaboration
by fitting MD's of samples of events with 2- and 3-jets
at different values of $\ymin$  are also shown for comparison in the last
column.}\label{table}
 \begin{center}
 \begin{tabular}{|c|c|c|c|c||c|c|}
\hline
& ALEPH  &   DELPHI &   OPAL  & SLD
& \multicolumn{2}{c|}{~~~DELPHI}    \\ \hline
& \multicolumn{4}{l||}{~~$\alpha$ = 0.463\hfill$\ymin$ = 0.02~~}
& \multicolumn{2}{l|}{$f$=0.463}  \\  \hline
$\bar n_1$ & 17.7$\pm$1.1 &  18.2$\pm$0.2     & 18.4$\pm$0.2 & 18.4$\pm$0.2 & $\bar
n_{2-jet}$ & 18.5$\pm$0.1\\
$k_1$       & 111$\pm$168 & 90$\pm$20  & 71$\pm$11 & 47$\pm$4 & $k_{2-jet}$ & 57$\pm$
4 \\
$\bar n_2$  &     23.6$\pm$0.8 & 23.9$\pm$0.2  &   24.0$\pm$0.2 &  23.0$\pm$0.2 &
$\bar
n_{3-jet}$ & 22.9$\pm$0.1 \\
$k_2$ &  32$\pm$15 & 31$\pm$3 &         28$\pm$2 &   29$\pm$2 &  $k_{3-jet}$
& 44$\pm$2\\
\chiDF & 3.56/22 & 8.95/21 &  3.32/21 & 17.6/21 & & \\  \hline
& \multicolumn{4}{l||}{~~$\alpha$ = 0.659\hfill$\ymin$ = 0.04~~}  &
\multicolumn{2}{l|}{$f$=0.659} \\  \hline
$\bar n_1$ & 18.5$\pm$0.7 & 18.9$\pm$0.2 & 19.0$\pm$0.1 & 18.9$\pm$0.1 & $\bar
n_{2-jet}$ &
19.4$\pm$0.1\\
$k_1$       & 66$\pm$46 & 63$\pm$8  & 54$\pm$5 & 42$\pm$3 & $k_{2-jet}$ & 44$\pm$2
\\
$\bar n_2$  &     25.5$\pm$1.0 & 25.8$\pm$0.3  &   25.9$\pm$0.2 &  24.7$\pm$0.2 &
$\bar
n_{3-jet}$ & 24.8$\pm$0.1 \\
$k_2$  &   47$\pm$33 &  44$\pm$5 & 40$\pm$5 &   37$\pm$3 &  $k_{3-jet}$ & 42$\pm$2\\
\chiDF & 3.72/22 & 10.1/21 &  4.40/21 & 16.3/21 & & \\  \hline
& \multicolumn{4}{l||}{~~$\alpha$ = 0.767\hfill$\ymin$ = 0.06~~}
& \multicolumn{2}{l|}{$f$=0.767} \\  \hline
$\bar n_1$ & 19.1$\pm$0.5 & 19.4$\pm$0.2 & 19.5$\pm$0.07 & 19.3$\pm$0.09 & $\bar
n_{2-jet}$ &
20.0$\pm$0.1\\
$k_1$       & 53$\pm$24 & 52$\pm$6  & 46$\pm$3 & 39$\pm$2 & $k_{2-jet}$ & 38$\pm$1
\\
$\bar n_2$  &     27.0$\pm$1.1 & 27.3$\pm$0.3  &   27.5$\pm$0.2 &  26.0$\pm$0.2 &
$\bar
n_{3-jet}$ & 26.0$\pm$0.1 \\
$k_2$  &   65$\pm$62 &  61$\pm$10 & 55$\pm$8 &   47$\pm$5 &  $k_{3-jet}$ & 45$\pm$2\\
\chiDF & 3.86/22 & 11.7/21 &  6.30/21 & 15.6/21 & & \\  \hline
& \multicolumn{4}{l||}{~~$\alpha$ = 0.834\hfill$\ymin$ = 0.08~~}
& \multicolumn{2}{l|}{$f$=0.834} \\  \hline
$\bar n_1$ & 19.5$\pm$0.4  &  19.8$\pm$0.1   & 19.9$\pm$0.6 & 19.6$\pm$0.1 & $\bar
n_{2-jet}$ & 20.4$\pm$0.1\\
$k_1$       & 45$\pm$15 & 46$\pm$3  & 40$\pm$2 & 37$\pm$2 & $k_{2-jet}$ & 34$\pm$1
\\
$\bar n_2$  &  28.2$\pm$1.2  & 28.6$\pm$0.3  &  28.8$\pm$0.2 &  27.1$\pm$0.3 &
$\bar
n_{3-jet}$ & 26.8$\pm$0.1 \\
$k_2$  &   92$\pm$121 &  85$\pm$18 &  76$\pm$15 &   59$\pm$7 &  $k_{3-jet}$ & 49$\pm$1 \\
\chiDF & 3.99/22 & 13.9/21 &  8.81/21 &  15.2/21 & & \\  \hline
 \end{tabular}
 \end{center}
\end{table}

\section*{Acknowledgments}
We would like to thank the organizers of this fruitful and very stimulating
workshop.
Our work was supported in part by M.U.R.S.T. (Italy) under grant 1995.

\end{document}